\documentstyle[sprocl]{article}
\newcommand{\matel}[3]{\langle #1|#2|#3\rangle}
\newcommand{\ra}{\rightarrow}

\newcommand{\sG}{\sigma \cdot G}

\newcommand{\MeV}{\,\mbox{MeV}}
 
\newcommand{\BR}{\,\mbox{branching ratio}}
\newcommand{\SL}{\,\mbox{semileptonic}}

\begin{document}
\begin{flushright}
UND-HEP-96-BIG01\\
hep-ph/??? \\
May 1996\\
\end{flushright}
\vspace{1cm}
\title{TREATING THE LIFETIMES OF CHARM AND BEAUTY HADRONS 
WITH QCD PLUS A BIT MORE! 
\footnote{Invited talk given at the Second Workshop on 
"Continuous Advances in QCD", Univ. of Minnesota, Minneapolis, March 1996.} } 

\author{I.I. Bigi }

\address{Physics Dept., University of Notre Dame du Lac, 
Notre Dame, IN 46556, U.S.A.\\
e-mail address: BIGI@UNDHEP.HEP.ND.EDU} 

\maketitle\abstracts{
The heavy quark expansion implemented through an operator 
product expansion provides us with a treatment of inclusive 
decays of beauty and charm hadrons that is genuinely derived 
from QCD,though it requires one additional assumption, namely that 
of 'local' quark-hadron duality. Subtleties in the application 
of factorization to hadronic expectation values are pointed out. 
The observed pattern in the charm lifetime ratios is reproduced 
in a semi-quantitative manner. The ratio 
$\tau (\Lambda _b)/\tau (B_d)$ cannot be pushed significantly 
below 0.9 -- unless one invokes a new hitherto unknown paradigm 
for evaluating baryonic matrix elements. One confidently predicts 
$\tau (B^-)$ to exceed $\tau (B_d)$ by several percent only. Failure 
of those predictions would force us to pay a hefty theoretical 
price, namely ultimately to abandon local 
duality as a practical concept.} 

The notion that the weak lifetimes of beauty and charm hadrons have to 
be {\em measured} accurately will hardly be challenged. For on the one hand the 
decay widths constitute a defining property of hadrons; on the other hand one 
has to know their size to translate the \SL \BR ~into a \SL ~width from which 
one can extract the KM parameters etc.; lastly 
a precise recording of the lifetime evolution is essential in probing 
$B^0 - \bar B^0$ oscillations as described by $\Delta m$ and 
$\Delta \Gamma$. 
The answer to "How well do we need to understand the 
lifetimes {\em theoretically}?" is however 
less obvious. For one can measure   
total widths accurately {\em without theoretical} input. Furthermore 
no apparent {\em qualitative} disaster has occurred since the expected pattern 
has indeed been observed, namely that the relative lifetime differences 
among beauty hadrons are smaller than among charm hadrons. Finally one can 
recite several reasons why a theoretical treatment of 
{\em nonleptonic} widths could -- or even 
should -- fail on a {\em quantitative} level. 
I, however, view the task of describing lifetimes of heavy flavour 
hadrons as a no-lose situation. For the general concept of quark-hadron 
duality implies that total decay widths represent the `safest' 
quantity theoretically after \SL~ widths. Once one has developed a 
description that is genuinely based on QCD, then even a failure of it 
will teach us a valuable -- albeit disappointing -- lesson on 
QCD, namely on quantitative limitations for the concept of duality. 
In Sect.1 I sketch the relevant methodology of the heavy quark 
expansion; in Sect.2 and 3 I compare the predictions on the 
widths of charm and beauty hadrons, respectively, with the 
available data before presenting a summary in Sect.4.      

\section{Methodology of the Heavy Quark Expansion}

The experimental findings in 1979 that the \SL \BR~ of $D^+$ mesons 
is much higher than that of $D^0$ mesons and that therefore the 
$D^+$ is much longer lived than the $D^0$ caused quite a stir in the 
community since it ran counter to some strongly held convictions. These  
data enforced revisions in our descriptions that first took shape in 
the form of phenomenological models: the concepts of 
Pauli Interference (PI)~\cite{PI}, 
Weak Annihilation (WA) \cite{SONI,MINKOWSKI} in meson  
and W Scattering (WS) \cite{BARYONS1,BARYONS2,BARYONS3} in baryon 
decays, respectively,  
were born. The foundations 
for a truly theoretical description of heavy flavour decays were laid 
already in 1983 by Shifman and Voloshin \cite{SV}: they argued that 
an expansion in powers of $1/m_Q$ with $m_Q$ being the heavy flavour quark 
mass can be performed for inclusive decay rates. The analysis involves a 
sequence of steps.  
In analogy to the treatment of 
$e^+e^-\rightarrow hadrons$ one first describes the transition rate into an 
inclusive final state $f$ through the imaginary part of a 
forward scattering operator evaluated to second order in the weak 
interactions \cite{SV,BUV,BS}:
$$\hat T(Q\rightarrow f\rightarrow Q)=
i \, {\rm Im}\, \int d^4x\{ {\cal L}_W(x){\cal L}_W^{\dagger}(0)\} 
_T\eqno(1)$$
$\{ .\} _T$ denotes the time ordered product and 
${\cal L}_W$ the relevant effective weak Lagrangian expressed on 
the 
parton level. If the energy released in the decay is sufficiently large 
one can express the {\em non-local} operator product in eq.(1) as an 
infinite sum of {\em local} operators $O_i$ of increasing dimension 
with 
coefficients $\tilde c_i$ 
containing higher and higher powers of  
$1/m_Q$ ~\footnote{It should be kept in mind, 
though, that 
it is primarily the {\em energy release} rather than $m_Q$ that controls the 
expansion.}. The width for $H_Q\rightarrow f$ is then obtained by 
taking the 
expectation value of $\hat T$ between the state $H_Q$:
$$\matel{H_Q}{\hat T (Q\ra f\ra Q)}{H_Q} \propto 
\Gamma (H_Q\ra f) = G_F^2 |KM|^2 
\sum _i \tilde c_i^{(f)}(\mu ) \matel{H_Q}{O_i}{H_Q}_{(\mu )} 
\eqno(2)$$
The c number coefficients $\tilde c_i^{(f)}(\mu )$ are determined by 
short-distance dynamics whereas long-distance dynamics controls the 
expectation values of the local operators $O_i$ \cite{MISHA}. 
Such a separation 
necessitates the introduction of an auxiliary scale with 
$ {\rm long \; distance} > \mu ^{-1} > {\rm short \; distance} $. 
While this is a conceptually and often also practically important 
point I will not refer to it explicitly anymore in this article  
~\footnote{Observables do not depend on $\mu$. Yet  
we have to choose $\Lambda _{QCD} \ll \mu \ll m_Q$ if we want 
to calculate perturbative as well as nonperturbative corrections in a 
self-consistent fashion.}. The coefficients $\tilde c^{(f)}(\mu )$ 
depend on the KM parameters and the quark masses; in particular, they 
contain powers of $1/m_Q$ that increase with the dimension of the 
local operators $O_i$. 

After a stumbling block concerning the $1/m_Q$ scaling in the presence 
of gluon radiation had been removed \cite{MIRAGE}  
this expansion was more 
fully performed in \cite{BUV,BS} with the following result: 
$$\Gamma (H_Q\ra f)=\frac{G_F^2m_Q^5}{192\pi ^3}|KM|^2
\left[ c_3^f\matel{H_Q}{\bar QQ}{H_Q}+
c_5^f\frac{
\matel{H_Q}{\bar Qi\sG Q}{H_Q}}{m_Q^2}+ \right.
$$ 
$$\left. +\sum _i c_{6,i}^f\frac{\matel{H_Q}
{(\bar Q\Gamma _iq)(\bar q\Gamma _iQ)}{H_Q}}
{m_Q^3} + {\cal O}(1/m_Q^4)\right]  \eqno(3)$$ 
with $KM$ denoting the product of the KM parameters. As 
already stated, the quantities $c_i^f$ can be calculated 
within short-distance dynamics; furthermore the operators 
appearing on the right hand side of eq.(3) are known and 
their dimensions control the scaling in $1/m_Q$ 
~\footnote{Contributions of order $1/m_Q^3$ arise also from 
expanding $\matel{H_Q}{\bar Qi \sigma \cdot GQ}{H_Q}/m_Q^2$; 
those are practically insensitive to the light quark flavours.}. 

Using the equations of motion one finds for the leading operator 
$\bar QQ$: 
$$\bar QQ= \bar Q\gamma _0Q -
\frac{\bar Q[(i\vec D)^2-(i/2)\sG]Q}{2m_Q^2}+
g_S^2\frac{\bar Q\gamma _0t^iQ
\sum _q \bar q\gamma _0 t^iq}{4m_Q^3} + 
{\cal O}(1/m_Q^4)\eqno(4)$$
with the sum in the last term running over the light quarks $q$; the 
$t^i$ denote the 
colour $SU(3)$ generators. Total derivatives are ignored in this 
expansion since they do not contribute to the expectation values.  
Since $\bar Q\gamma _0 Q$ constitutes the Noether 
current for the heavy-flavour quantum number one has 
$\matel{H_Q}{\bar Q\gamma _0Q}{H_Q}_{norm}=1$ 
\footnote{The relativistic normalization is used: 
$\matel{H_Q}{O_i}{H_Q}_{norm} \equiv 
\matel{H_Q}{O_i}{H_Q}/2M_{H_Q}$.} 
leading to   
$$\matel{H_Q}{\bar QQ}{H_Q}_{norm}=1+ 
{\cal O}(1/m_Q^2) \eqno(5)$$ 
From eqs.(3) and (5) we read off two important general results: 
\begin{itemize}
\item The naive spectator contribution 
$\Gamma _{spect}(H_Q)\propto G_F^2 m_Q^5$ 
emerges from $\matel{H_Q}{\bar QQ}{H_Q}$ as the leading term for 
$m_Q \ra \infty$. 

\item {\em There is no 
pre-asymptotic correction of order $1/m_Q$!} For the only 
{\em locally} gauge invariant operator of dimension-four -- 
$\bar Q i \gamma _{\mu}D_{\mu}Q,\, D_{\mu}= \partial _{\mu} 
-ig_S A_{\mu}^i t^i$ -- can be reduced to $m_Q \bar QQ$ due 
to the equation of motion. This yields the general result that the 
leading non-perturbative corrections to beauty decays are of 
order $(\mu _{had}/m_b)^2 \sim 
{\cal O}\left( (1\, {\rm GeV}/m_b)^2\right) 
\sim$ few \%, i.e., quite small. Among other things this implies that 
$B_c$ decays exhibit a short lifetime below 1 psec and that their 
decays are dominated by charm decays \cite{BCLAB,BENEKE,LUSIGNOLI}. 

\item 
Two dimension-five operators emerge, namely 
$\bar Q (i\vec D)^2 Q$ and $\bar Q \sG Q$, which had been overlooked  
in the phenomenological approaches. The first one represents the 
square of the spatial momentum of the heavy quark $Q$ moving in 
the soft gluon background and thus describes its kinetic 
energy \footnote{Since it is not a Lorentz scalar, it cannot 
appear in 
eq.(3).}. The second one constitutes the chromomagnetic operator. 

\item PI, WA and WS that had been anticipated 
in the phenomenological descriptions enter through  
$\matel{H_Q}{(\bar Q\Gamma _iq)(\bar q\Gamma _iQ)}{H_Q}$ in order 
$1/m_Q^3$. The formally leading contributions to WA are 
helicity suppressed \cite{MIRAGE,WA}.   

\end{itemize}
\noindent These points can be summarized as follows:
$$ \Gamma (H_Q) = 
\Gamma _{decay}(H_Q) + \Gamma _{PI,WA,WS}(H_Q) + 
{\cal O}(1/m_Q^4) $$
$$
\Gamma _{decay}(H_Q) = \Gamma _{spect}(H_Q) + {\cal O}( 1/m_Q^2) 
\eqno(6) $$
$\Gamma _{spect}$ is universal for all hadrons of a given flavour, 
but $\Gamma _{decay}$ is not:   
$ \Gamma _{decay}(P_Q) \neq \Gamma _{decay}(\Lambda _Q) 
\left( \neq \Gamma _{decay}(\Omega _Q)\right) $. 

The mesonic matrix elements of the 
chromomagnetic operator 
can be extracted from the hyperfine splitting: 
$$\langle \mu _G^2\rangle _{H_Q} \equiv 
\matel{P_Q}{\bar Q\frac{i}{2}\sG Q}{P_Q}_{norm} \simeq 
\frac{3}{2} m_Q (M_{V_Q}-M_{P_Q}) \simeq 
\frac{3}{4} (M_{V_Q}^2-M_{P_Q}^2) \eqno(7a)$$
where $V_Q = B^*,\, D^*$ and $P_Q = B,\, D$ 
~\footnote{We have also assumed here that the mass of the 
antiquark in the meson is light and can be neglected to this order: 
$m_Q \simeq (M_{V_Q} + M_{P_Q}) /2$. For $B_c$ mesons one obviously 
has to go beyond this approximation.}. Thus 
$$\langle \mu _G^2\rangle _D \simeq 0.41\, {\rm (GeV)^2}\; , \;  
\langle \mu _G^2\rangle _B \simeq 0.37\, {\rm (GeV)^2}\; , \; 
\frac{\langle \mu _G^2\rangle _D}{m_c^2} \simeq 0.21 \; , \;  
\frac{\langle \mu _G^2\rangle _B}{m_b^2} \simeq 0.016 \eqno(7b) 
$$
A measure for the numerical reliability of the expansion is then 
provided by $\sqrt{\langle \mu _G^2\rangle _D/m_c^2} \simeq 0.46$ 
and $\sqrt{\langle \mu _G^2\rangle _B/m_b^2} \simeq 0.13$, 
respectively. This parameter is certainly small compared to unity 
for beauty decays; on the other hand a $1/m_c$ 
expansion is of uncertain numerical value. 

\noindent The light di-quark system in $\Lambda _Q$ and 
$\Xi _Q$ baryons carries no spin; therefore   
$$\matel{\Lambda _Q}{\bar Q \sigma \cdot GQ}{\Lambda _Q} \simeq 0 
\simeq \matel{\Xi _Q}{\bar Q \sigma \cdot GQ}{\Xi _Q}\eqno(8) 
$$
This operator thus generates width differences between mesons 
and baryons in order $1/m_Q^2$, see eqs.(7b) vs. (8). 

The value of $\matel{H_Q}{\bar Q(i\vec D)^2Q}{H_Q}_{norm}
\equiv \langle (\vec p_Q)^2\rangle _{H_Q}$ is not known accurately. 
An analysis based on QCD sum rules yields \cite{QCDSR}  
$$  \langle (\vec p_b)^2\rangle _B \simeq 0.5 \pm 0.1\; {\rm (GeV)^2} 
\eqno(9)
$$ 
in agreement with a rigorous lower bound   
\cite{VOLOSHIN,OPTICAL} 
$$\langle (\vec p_b)^2\rangle _B \geq \langle \mu _G^2\rangle _B $$
The differences in the mesonic and baryonic expectation values 
can be related to the `spin averaged' meson and baryon masses: 
$ \langle (\vec p_Q)^2\rangle _{\Lambda _Q}-
\langle (\vec p_Q)^2\rangle _{P_Q} \simeq 
\frac{2m_bm_c}{m_b-m_c}\cdot 
\{ [\langle M_D\rangle -M_{\Lambda _c}]- 
[\langle M_B\rangle -M_{\Lambda _b}] \}$ \cite{BUVPREPRINT}. 
Present data yield:  
$$ \langle (\vec p_Q)^2\rangle _{\Lambda _Q}-
\langle (\vec p_Q)^2\rangle _{P_Q} =  
- (0.015 \pm 0.030)\; {\rm (GeV)^2} \eqno(10)$$
i.e., no significant difference.  
In deriving eq.(10) it was assumed that 
the $c$ quark can be treated as heavy; in that case  
$\langle (\vec p_c)^2\rangle _{H_c} \simeq 
\langle (\vec p_b)^2\rangle _{H_b}$ holds. 

The expectation values of the four-fermion operators 
are not reliably known. To estimate their size for {\em mesons} 
one usually invokes {\em factorization}:   
$$\matel{H_Q(p)}{(\bar Q_L\gamma _{\mu}q_L)
(\bar q_L\gamma _{\nu}Q_L}{H_Q(p)}_{norm}\simeq$$
$$
\matel{H_Q(p)}{(\bar Q_L\gamma _{\mu}q_L)}{0}_{norm} 
\matel{0}{(\bar q_L\gamma _{\nu}Q_L}{H_Q(p)}_{norm} = 
\frac{1}{8 M_{H_Q}}f^2_{H_Q}p_{\mu}p_{\nu}\eqno(11a)$$   
$$\matel{H_Q(p)}{(\bar Q_L\gamma _{\mu}\lambda _iq_L)
(\bar q_L\gamma _{\nu}\lambda _iQ_L}{H_Q(p)}_{norm}\simeq 
$$
$$\matel{H_Q(p)}{(\bar Q_L\gamma _{\mu}\lambda _i q_L)}{0}_{norm} 
\matel{0}{(\bar q_L\gamma _{\nu}\lambda _i Q_L}{H_Q(p)}_{norm} = 0 
\eqno(11b)$$   
However such an ansatz cannot be an identity. It can hold as an approximation, 
but only for certain scales. Invoking it at $\sim m_Q$ does {\em not} 
make sense at all. For as far as QCD is concerned, $m_Q$ is a completely 
foreign quantity, only moderately less so than the mass of an 
elephant. A priori it has a chance to hold at ordinary hadronic scales 
$\mu _{had} \sim 0.5 \div 1$ GeV \cite{HYBRID}; 
various theoretical analyses based on 
QCD sum rules, QCD lattice simulations, $1/N_C$ expansions etc. have 
indeed found it to apply in that regime. It would be  
inadequate conceptually as well as numerically to renormalize 
merely the decay constant: $f_Q(m_Q) \ra f_Q(\mu _{had})$. Instead the 
full set of operators has to be evaluated at $\mu _{had}$. One proceeds 
in three steps (for details see \cite{BELLINI}): 

\noindent (A) Ultraviolet renormalization translates 
the weak Lagrangian defined at $M_W$, ${\cal L}_W(M_W)$, into 
one effective at $m_Q$, ${\cal L}_W(m_Q)$. 

\noindent (B) All operators ${\cal O}_i$ in eq.(2) undergo hybrid 
renormalization \cite{HYBRID} down to $\mu _{had}$. 

\noindent (C) At scale $\mu _{had}$ one invokes factorization. 

Some comments are in order to elucidate the situation that is 
not properly reflected in \cite{NS}: 
\begin{itemize} 
\item It has been known for more than 16 years now that the 
factorizable contributions to PI almost cancel -- 
apparently for accidental reasons -- at scales around $m_Q$ 
making the ratio of non-factorizable to factorizable contributions 
large and numerically unstable there. 
\item No such cancellation occurs around scales $\mu _{had}$ making  
factorizable contributions numerically stable and dominant over 
non-factorizable ones. 
\item Contributions that are factorizable (in colour space) 
at $\mu _{had}$ are mainly 
non-factorizable at $m_Q$. 
\item The role of 
non-factorizable terms has been addressed in the literature over the 
years, most explicitely and in a most detailed way in \cite{DS,WA}. 
\end{itemize}

The situation becomes much more complex for  
baryon decays. To order $1/m_Q^3$ there are several different 
ways in which the valence quarks of the baryon can be contracted 
with the quark fields in the four-quark operators; furthermore 
WS is {\em not} helicity suppressed and thus can make a sizeable 
contribution to lifetime differences; also the PI effects can 
now be constructive 
as well as destructive. Finally one cannot take 
recourse to factorisation as a limiting case.  
Thus there emerge three types of 
numerically significant mechanisms at this order in baryon 
decays -- in contrast to meson decays where there is a 
{\em single}  dominant 
source for lifetime differences -- and their strength cannot be 
expressed in terms of 
a single observable like $f_{H_Q}$. At present we do not know 
how to determine the relevant matrix elements in a 
model-independant way. The best available guidance and 
inspiration is to be derived from quark model calculations with 
their inherent uncertainties. This analysis had already been 
undertaken 
in the framework of phenomenological models 
\cite{BARYONS1,BARYONS2,BARYONS3}. 
One thing should be obvious already at this point: 
with terms of different signs and somewhat uncertain size 
contributing to differences among baryon lifetimes one has to 
take even semi-quantitative predictions with 
a grain of salt!

\section{Lifetimes of Charm Hadrons -- `A Painting in Broad 
Brush Strokes'} 

In discussing heavy quark expansions one should start with three 
caveats: (i) Since the charm quark mass is not much larger than 
hadronic scales, the expansion parameter is uncomfortably large, 
though smaller than unity: $\mu _{had} /m_c \sim 0.5$. (ii) By the 
same token the evaluation of hybrid renormalization that 
turns out to be quantitatively important is of uncertain  
numerical reliability. (iii) Equating the observed semileptonic 
width of D mesons with the theoretical expression through order 
$1/m_c^2$ yields $m_c \simeq 1.6$ GeV. The theoretically more 
reasonable value $m_c \simeq 1.4$ GeV reproduces only half of 
$\Gamma _{SL}(D)|_{exp}$. Terms of order $1/m_c^3$ in 
$\Gamma _{SL}(D)|_{theor.}$ do not seem to bridge the 
gap \cite{DIKEMAN}. This 
discrepancy can be interpreted as signaling that quark-hadron duality 
does not generally hold even in \SL~ charm decays. I will adopt the 
working hypothesis that it still applies -- with reasonable 
accuracy -- to the {\em ratios} of lifetimes and \SL~ branching ratios.  
With these caveats one can dare to make 
predictions on the charm lifetime ratios. 

The dominant source for the $D^+$-$D^0$ lifetime difference 
is destructive PI in nonleptonic $D^+$ decays with WA enhancing 
the $D^0$ width as a secondary effect. More specifically one 
finds \cite{MIRAGE}  
$$\Gamma (D^+)\simeq \Gamma _{decay}(D)+
\Gamma _{PI}(D^+)\eqno(12a)$$
$$ \Gamma _{PI}(D^+)\simeq \Gamma _0\cdot 
24\pi ^2\frac{f_D^2}{m_c^2}\kappa ^{-4} 
\left[ (c_+^2-c_-^2)\kappa ^{\frac{9}{2}}+
\frac{c_+^2+c_-^2}{3} - 
\frac{1}{9}(\kappa ^{\frac{9}{2}}-1)(c_+^2-c_-^2)
\right] \; , 
\eqno(12b) $$
where 
$\kappa \equiv [\alpha _S(\mu _{had})/\alpha _S(m_c)]^{1/b}$, 
$b = 11 - 2n_F/3$ 
represents hybrid renormalization. 
Large cancellations no longer occur among the 
factorizable terms in Eqs.(12); their 
overall contribution is destructive 
and large. We then arrive at 
$$ \frac{\tau (D^+)}{\tau (D^0)} \simeq 1 +  
\left( \frac{f_D}{200\, {\rm MeV}}\right) ^2 \sim 2 \eqno(13) $$ 
A priori $\tau (D_s)$ and $\tau (D^0)$ 
could differ substantially from each other, 
in particular due to a different weight of WA in the two 
transitions. Yet using the heavy quark expansion 
and assuming factorization one predicts $\tau (D_s)$ and 
$\tau (D^0)$ to agree within several percent \cite{DS} 
due to a compensation among various competing smallish effects. 

The main differences in the lifetimes of baryons on one hand and of 
mesons on the other and also among the various baryons arise in 
order $1/m_c^3$ due to WS and destructive as well as 
constructive PI \cite{BARYONS1,BARYONS2,BARYONS3}: 
$$\Gamma (\Lambda _c^+)= \Gamma _{decay}(\Lambda _c^+) +
\Gamma _{WS}(\Lambda _c^+) -
|\Gamma _{PI,-}(\Lambda _c)|   \eqno(14a)$$
$$\Gamma (\Xi _c^0)= \Gamma _{decay}(\Xi _c^0) +
\Gamma _{WS}(\Xi _c^0) +
|\Gamma _{PI,+}(\Xi _c^0)|   \eqno(14b)$$
$$\Gamma (\Xi _c^+)= \Gamma _{decay}(\Xi _c^+) +
|\Gamma _{PI,+}(\Xi _c^+)| -
|\Gamma _{PI,-}(\Xi _c^+)|   \eqno(14c)$$
$$\Gamma (\Omega _c) = \Gamma _{decay}(\Omega _c) + 
|\Gamma _{PI,+}(\Omega _c)| \eqno(14d)$$
with both quantities on the right-hand-side of eq.(14d) 
differing from the corresponding ones for $\Lambda _c$ 
or $\Xi _c$ decays. 
On rather general grounds one concludes:
$$\tau (\Xi _c^0) < \tau (\Xi _c^+)\; , \; \; \; 
\tau (\Xi _c^0) < \tau (\Lambda _c^+) \eqno(15)$$
To go beyond this qualitative prediction one has to 
evaluate the expectation values of the various 
four-fermion operators. No model-independant manner is 
known for doing that for baryons; we do not even have a concept like 
factorization allowing us to lump our ignorence into a single 
quantity. Instead we have to rely on quark model 
computations and thus have to be prepared for 
additional very sizeable theoretical uncertainties. 
In Table \ref{TABLE10} I juxtapose the data with the 
theoretical expectations obtained from the 
heavy quark expansion described
above. The numbers for baryon lifetimes are based on quark model 
evaluations of the four-fermion expectation values; this is 
indicated by an asterisk.   
Details can be found in \cite{BELLINI}.  
\begin{table}[t] 
\caption{QCD Predictions for Charm Lifetime Ratios  
\label{TABLE10}}
\vspace{0.4cm}
\begin{center}   
\begin{tabular} {|l|l|l|l|}
\hline
Observable &QCD Expectations ($1/m_c$ expansion) & Ref. &
Data from \cite{BELLINI} \\ 
\hline 
\hline 
$\tau (D^+)/\tau (D^0)$ & $\sim 2 \; \; \; $ 
[for $f_D \simeq 200$ MeV] & \cite{MIRAGE} & $2.547 \pm 0.043$ \\ 
&(mainly due to {\em destructive} interference)& & \\ 
\hline 
$\tau (D_s)/\tau (D^0)$ &$1\pm$ few $\times 0.01$   
& \cite{DS} &  $ 1.12\pm 0.04$ \\
\hline 
$\tau (\Lambda _c)/\tau (D^0)$&$\sim 0.5 ^*$  & \cite{MARBELLA} & 
$0.51\pm 0.05$\\
\hline 
$\tau (\Xi ^+ _c)/\tau (\Lambda _c)$&$\sim 1.3 ^*$ & \cite{MARBELLA} &  
$1.75\pm 0.36$\\
\hline
$\tau (\Xi ^+ _c)/\tau (\Xi ^0 _c)$&$\sim 2.8 ^*$ & \cite{MARBELLA} & 
$3.57\pm 0.91$\\
\hline 
$\tau (\Xi ^+ _c)/\tau (\Omega _c)$&$\sim 4 ^*$ & \cite{MARBELLA} &  
$3.9 \pm 1.7$\\
\hline
\end{tabular}
\end{center} 
\end{table} 
 
The agreement between the expectations and the data, within the 
uncertainties,  is respectable or even 
remarkable considering the large theoretical expansion parameter 
and the fact that the lifetimes for the apparently shortest-lived 
hadron -- $\Omega _c$ -- and for the longest-lived one -- 
$D^+$ -- differ by an order of magnitude! Of course the experimental 
uncertainties in $\tau (\Xi _c)$ and $\tau (\Omega _c)$ are still 
large; the present agreement could fade away -- or even evaporate -- 
with the advent of more accurate data. Yet at present we conclude: 
\begin{itemize} 
\item The observed difference in $\tau (D^0)$ vs. $\tau (D^+)$ is 
understood as due mainly, though not exclusively, to a destructive 
interference in $\Gamma _{NL}(D^+)$ arising in order $1/m_c^3$. This is 
{\em not} contradicted by the data showing $BR_{SL}(D^+)\simeq 17$ \%. 
For the corrections of order $1/m_c^2$ reduce the number obtained in the 
naive spectator model -- $BR_{SL}(D)\simeq BR_{SL}(c)$ -- from around 
16\% down to around 9\% \cite{BUV}! 

\item The observed near-equality of $\tau (D^0)$ and $\tau (D_s)$ 
provides us with circumstantial evidence for the reduced weight 
of WA. It puts a severe bound on the size of the 
non-factorizable parts in the expectation values of 
the four-fermion operators, as given in \cite{DS}.  

\item The lifetimes of the charm baryons reflect the interplay of 
destructive as well as constructive PI and WS. 

\item The $\Omega _c$ naturally emerges as the shortest-lived charm 
hadron due to spin-spin interactions between the decaying $c$ 
quark and the spin-one $ss$ di-quark system. 

\end{itemize}  

\noindent Finally one should note that the ratios 
$\Gamma _{SL}(\Xi _c)/\Gamma _{SL}(D^0)$ and 
$\Gamma _{SL}(\Omega _c)/\Gamma _{SL}(D^0)$ will {\em not} reflect 
their lifetime ratios; for $\Gamma _{SL}(\Xi _c)$ and 
$\Gamma _{SL}(\Omega _c)$ get significantly enhanced relative to 
$\Gamma _{SL}(D^0)$ in order $1/m_c^3$ due to {\em constructive} PI in 
$\Gamma _{SL}(\Xi _c, \Omega _c)$ among the $s$ quarks 
\cite{VOLOSHIN2}. Thus $\Omega _c$ 
-- despite its short lifetime -- could well exhibit a larger 
\SL \BR~ than $D^0$!

\section{Lifetimes of Beauty Hadrons -- `Hic Rhodus, Hic Salta!'}

Most of the caveats stated for charm decays cannot be used as excuses 
for failures in beauty decays. Due to $m_b \gg \mu _{had}$ the heavy quark 
expansion would be expected to yield fairly reliable predictions on lifetime 
ratios among beauty hadrons. 
The actual computations proceed in close analogy to the charm case 
and can be found in \cite{BELLINI}.  
The $B_d - B^-$ lifetime difference is again driven mainly by 
destructive PI, namely in the $b\ra c \bar ud$ channel; similarly,  
$\tau (\Lambda _b)$ is reduced relative to $\tau (B_d)$ by 
WS winning out over destructive PI in $b\ra c \bar ud$: 
$$\Gamma (B_d) \simeq \Gamma _{decay}(B_d) \; , \; 
\Gamma (\Lambda _b) \simeq \Gamma _{decay}(\Lambda _b) 
+ \Gamma _{WS}(\Lambda _b) - |\Gamma _{PI,-}(\Lambda _b) |
$$
In Table \ref{TABLE2} 
I list presently available data 
together with quantitative predictions. 

\begin{table}[t]
\caption{QCD Predictions for Beauty Lifetimes  
\label{TABLE2}}  
\begin{center}
\begin{tabular} {|l|l|l|l|}
\hline
Observable &QCD Expectations ($1/m_b$ expansion)& Ref. &
Data from \cite{BELLINI}\\ 
\hline 
\hline 
$\tau (B^-)/\tau (B_d)$ & $1+
0.05(f_B/200\, \MeV )^2
[1\pm {\cal O}(30\%)]>1$ & \cite{MIRAGE} & $1.03 \pm 0.06$ \\
&(mainly due to {\em destructive} interference)&  & \\ 
\hline 
$\bar \tau (B_s)/\tau (B_d)$ &$1\pm {\cal O}(0.01)$ & \cite{STONE2}  
&  $ 0.97\pm 0.08$ \\
\hline 
$\tau (\Lambda _b)/\tau (B_d)$&$\simeq 0.9 ^*$ & \cite{STONE2} & 
$0.73\pm 0.06$ \\
\hline 
\end{tabular}
\end{center} 
\end{table} 

Several comments are in order here: 

\begin {itemize} 

\item These are predictions in the old-fashioned 
sense, i.e. they were made before data (or data of comparable 
sensitivity) became available. 

\item As far as the meson lifetimes are concerned, 
data and predictions are completely and {\em non-trivially} 
consistent. 
 
\item A careful evaluation of the radiative corrections and 
analysis of non-factorizable contributions allows to 
predict that $\tau (B^-)$ {\em exceeds} 
$\tau (B_d)$ by several percent, as stated in  
Table \ref{TABLE2}. Contrary to the claims of  
\cite{NS} future experimental findings that 
$\tau (B^-) < \tau (B_d)$ or 
$\tau (B^-) \simeq 1.2 \cdot  \tau (B_d)$ could {\em not naturally} be 
accommodated within the heavy quark expansion. One more cross 
check can be performed to make this case conclusive by closing a 
possible loophole in the argument: 
contrary to presently available theoretical evidence 
{\em factorization} might be a poor ansatz. One can extract 
the factorizable as well as non-factorizable contributions 
from a difference observed in the endpoint energy spectra 
for \SL~ decays of $B_d$ and $B^-$ mesons \cite{WA}. 
Comparing 
the inclusive lepton spectra in \SL $D^0$, $D^+$ and $D_s$ decays 
would provide us with similar information. 

\item The average $B_s$ lifetime, i.e. $\bar \tau (B_s) = 
[\tau (B_{s,{\rm long}}) + \tau (B_{s,{\rm short}})]/2$, 
as measured in $B_s \ra l \nu D_s^{(*)}$, is practically 
idential to $\tau (B_d)$. 

\item The largest lifetime difference among beauty {\rm mesons} is 
expected to occur due to $B_s - \bar B_s$ oscillations. One predicts 
\cite{BSBS}: 
$$
\frac{\Delta \Gamma (B_s)}{\bar \Gamma (B_s)} \equiv 
\frac{\Gamma (B_{s,{\rm short}}) - \Gamma (B_{s,{\rm long }})}
{\bar \Gamma (B_s)} \simeq 0.18 \cdot 
\frac{(f_{B_s})^2}{(200 \, {\rm MeV})^2} 
\eqno(16)
$$ 

\item The prediction on $\tau (\Lambda _b)/\tau (B_d)$ seems to be 
in conflict with the data. 

\end{itemize}
For proper evaluation of the last point one has to keep the following 
in mind: 

\noindent (i) The experimental situation has not been settled yet. 
In Table \ref{TABLE2}  I have listed the world average of 
{\em already published} data on $\tau (\Lambda _b)/\tau (B_d)$. It should 
be pointed out that a recent preliminary CDF study finds 
$\tau (\Lambda _b)/\tau (B_d) = 0.85 \pm 0.12$. While this value is 
quite consistent with the stated world average, it would also satisfy the 
theoretical prediction. 

\noindent (ii) The difference between 
$\tau (\Lambda _b)/\tau (B_d)|_{exp} \simeq 0.73$ and 
$\tau (\Lambda _b)/\tau (B_d)|_{theor} \simeq 0.9$ represents a 
large discrepancy. For once one has established -- as we have -- 
that $\tau (\Lambda _b)$ and $\tau (B_d)$ have to coincide for 
$m_b \ra \infty$, then the predictions really concern the 
deviation from unity; finding a $\sim 27$ \% deviation when one 
around 10 \% was predicted amounts to an error of order 
300 \%! 

\noindent (iii) A failure of that proportion cannot be rectified 
unless one adopts a new paradigm in evaluating baryonic expectation values. 
Two recent papers \cite{BOOST,NS} have re-analyzed the relevant 
{\em quark model} calculations and found: :   
$$ \tau (\Lambda _b)/\tau (B_d) \equiv 1 - {\rm DEV}\, , \; 
{\rm DEV} \sim 0.03 \div 0.12 \eqno(17)$$ 
i.e., indeed there are large  theoretical uncertainties in DEV, yet 
one cannot boost its size much beyond the 10 \% level. To achieve the 
latter one had to go {\em beyond} a description of baryons in terms of three 
valence quarks only.

\section{Summary}
 
During the last few years considerable conceptual progress has been 
achieved in our theoretical description of the decays of heavy flavour 
hadrons in general and their lifetimes in particular. Questions that could 
hardly be raised before can be tackled now. A failure to describe the weak 
lifetimes will of course never rule out QCD -- yet it can and will 
teach us significant lessons on the inner workings of QCD. 

Such failures can actually occur at different levels thus 
leading to different layers of lessons which I am going to list now in 
ascending order of depth: 
 
\noindent (i) It seems quite unlikely that future data could contradict 
the predicted qualitative pattern, namely $\tau (D^+) > 
\tau (D^0) \simeq \tau (D_s) > \tau (\Lambda _c)$ and 
$\tau (\Xi _c^0) < \tau (\Xi _c^+)$, 
$\tau (\Xi _c^0) < \tau (\Lambda _c)$.  

\noindent (ii) An inability to quantitatively 
reproduce the observed lifetime 
ratios for {\em charm baryons} can be rationalized most easily. For  
their widths receive contributions with different signs 
from several mechanisms whose intervention reflects the rather 
complex internal structure of baryons; furthermore contributions that 
are formally of higher order in $1/m_c$ are numerically 
reduced only in a moderate fashion; it would 
therefore seem unrealistic to 
expect any success beyond purely qualitative considerations. 

\noindent (iii) If however one 
succeeds in describing charm baryon 
lifetime ratios in a semi-quantitative fashion at least, then 
one can use this information to probe the internal structure of these 
baryons in a novel way, namely concerning the behaviour of the diquark 
system, as briefly referred to above for $\Omega _c$. 

\noindent (iv) If a future determination of 
$f_D$ revealed a significant 
discrepancy in the prediction for $\tau (D^+)/\tau (D^0)$, one could blame 
that on $m_c$ being too low. More specifically it would -- like 
$\Gamma _{SL}(D)$ signal the limitation of 
quark-hadron duality at the relatively low scale $m_c$. 

\noindent (v) For inclusive beauty decays no plausible deniability 
exists and one had to face up to harder lessons. 

\noindent (vi) As discussed before one has to allow for considerable 
numerical uncertainties in the predictions on 
$\tau (\Xi _b^0)$ vs. $\tau (\Xi _b^-)$ vs. 
$\tau (\Lambda _b)$ vs. $\tau (B_d)$. Yet their differences should 
not exceed the 10 \% level. To reproduce larger lifetime differences -- 
as suggested by the present world average on 
$\tau (\Lambda _b)/\tau (B_d)$ -- would require a new paradigm 
in evaluating at least baryonic matrix elements that goes beyond the 
usual valence quark description. Instead one might go one step further 
and argue that quark-hadron duality does not operate here with sufficient 
accuracy. 

\noindent (vii) Discrepancies concerning $B$ meson 
lifetime ratios would lead to 
unequivocal lessons.  Namely a failure in 
$\tau (B^-)/\tau (B_d)$ would first cast serious doubts on the applicability 
of factorization even at the natural {\em low} scale. 
However if an extraction of the 
expectation values of the four-fermion operators from \SL ~decays 
without imposing factorization had closed this loophole, one would 
be forced to conclude that local duality is not realized in 
nonleptonic beauty decays; {\em local} duality means that the rates for 
inclusive processes involving hadrons can be calculated from the 
corresponding quark reactions  
{\em without} the 'smearing' or averaging in energy advocated in 
\cite{POGGIO}. The same negative conclusion would follow if 
$\bar \tau (B_s)$ and $\tau (B_d)$ were found to differ by more than 
a few percent.  
This would certainly be a disappointing lesson -- 
in particular since we have not spotted any previous sign for trouble -- 
but it would be an important one nevertheless! A more detailed 
discussion of these points can be found in \cite{MISHA,BELLINI}.

\section*{Acknowledgements} 
I thoroughly enjoyed the lively and inspiring 
atmosphere created by the organizers of this smoothly running 
meeting. This work was supported by the National Science Foundation under 
grant number PHY 92-13313.


\begin{thebibliography}{99}

\bibitem{PI} 
B.Guberina, S.Nussinov, R.Peccei, R.R\" uckl, 
{\em Phys.Lett.} {\bf B89} (1979) 111.

\bibitem{SONI} 
M. Bander, D. Silverman, A. Soni, 
{\em Phys. Rev. Lett.} {\bf 44} (1980) 7.

\bibitem{MINKOWSKI} 
H.Fritzsch, P.Minkowski, 
{\em Phys.Lett.} {\bf B90} (1980) 455; 
W.Bernreuther et al., 
{\em Z.Phys.} {\bf C4} (1980) 257; 
I.I.Bigi, {\em Z.Phys.} {\bf C5} (1980) 313. 

\bibitem{BARYONS1} 
N. Bilic, B. Guberina, J. Trampetic, 
{\em Nucl. Phys.} {\bf B248} (1984) 261; 
M. Shifman, M. Voloshin, {\em Sov. J. Nucl. Phys.} 
{\bf 41} (1985) 120. 

\bibitem{BARYONS2} 
M. Shifman, M. Voloshin, {\em JETP} 
{\bf 64} (1986) 698. 

\bibitem{BARYONS3} 
B. Guberina, R. R\" uckl, J. Trampetic, 
{\em Z. Phys.} {\bf C33} (1986) 297. 

\bibitem{SV}
M.Shifman, M.Voloshin, 1982, in: V.Khoze, M.Shifman, 
{\em Sov.Phys.Uspekhi} (1983) 387; {\em Sov.Journ.Nucl.Phys.} 
{\bf 41} (1985) 120. 

\bibitem{BUV} 
I.I. Bigi, N.G. Uraltsev, A. Vainshtein, {\em Phys. Lett.} {\bf B293} 
(1992) 430; (E) {\bf B297} (1993) 477.

\bibitem{BS}
B. Blok, M. Shifman, {\em Nucl. Phys.} {\bf B399} (1993) 441; 459.

\bibitem{MISHA}
M. Shifman, talk given 
at the V Intern. Symp. on Particles, Strings 
and Cosmology -- PASCOS --, John Hopkins Univ., Baltimore, 
March 1995, to appear in the Proceed., preprint 
TPI-MINN-95/15-T [hep-ph/9505289]. 

\bibitem{MIRAGE}
I.I. Bigi, N.G. Uraltsev, {\em Phys. Lett.} {\bf B280} (1992) 120. 

\bibitem{BCLAB}
I.I.Bigi, {\em Acta Phys.Polon.} {\bf B26} (1995) 641;   
{\em Phys.Lett.} {\bf B371} (1996) 105. 

\bibitem{BENEKE}
M. Beneke, G. Buchalla , {\em Phys.Rev.} {\bf D53} (1996) 4991.  

\bibitem{LUSIGNOLI}
M. Lusignoli, M.Masetti, {\em Z.Phys.} {\bf C51} (1991) 549. 

\bibitem{QCDSR} 
This is an update of: 
P. Ball, V. Braun, {\em Phys. Rev.} {\bf D49} (1994) 2472.


\bibitem{VOLOSHIN} 
M. Voloshin,  {\em Surv. High En. Phys.} {\bf 8} (1995) 27; 
this analysis improved 
upon an inequality previously derived in: I.I. Bigi, M. Shifman, N.G. 
Uraltsev, A. Vainshtein, {\em Int. Journ. Mod. Phys. A} {\bf 9} (1994) 
2467. 

\bibitem{OPTICAL} 
I.I.Bigi et al., {\em Phys.Rev.} {\bf D52} (1995) 196. 

\bibitem{BUVPREPRINT}
I.I. Bigi, N.G. Uraltsev, A. Vainshtein, preprint FERMILAB-PUB-
92/158-T; 
due to length limitations this formula was left out in the journal 
version, ref.1; I.I. Bigi, in: Proc. of the XXVI Intern. Conf. on 
High Energy Physics, Dallas, 1992, J.R. Sanford (ed.), AIP, 
No. 272, p. 402.   

\bibitem{HYBRID} 
M. Shifman, M. Voloshin, {\em JETP} {\bf 64} (1986) 698; 
{\em Sov. J. Nucl. Phys.} {\bf 45} (1987) 292; 
H. Politzer, M. Wise, {\em Phys. Lett.} {\bf B206} (1988) 681. 

\bibitem{NS}
M. Neubert, C.T. Sachrajda, preprint CERN-TH/96-19 (1996). 

\bibitem{DS}
I.I. Bigi, N.G. Uraltsev, {\em Z. Physik} {\bf C62} (1994) 623. 

\bibitem{WA} 
I.I. Bigi, N.G. Uraltsev, {\em Nucl. Phys.} {\bf B423} (1994) 33. 

\bibitem{DIKEMAN}
B. Blok, R. Dikeman, M. Shifman, {\em Phys. Rev.} 
{\bf D51} (1995) 6167. 

\bibitem{MARBELLA}
B. Blok, M. Shifman, in: Proceedings of the Third Workshop on the 
Physics at a Tau-Charm Factory, Marbella, Spain, June 1993, R. \& J. 
Kirkby (eds.), Editions Frontieres, 1994. 

\bibitem{BELLINI} 
for a comprehensive review, see: G. Bellini, I.I. Bigi, P. Dornan, 
{\bf Physics Rep.}, to appear. 

\bibitem{VOLOSHIN2} 
M. Voloshin, preprint TPI-MINN-96/4-T, hep-ph/9604335.  

\bibitem{BOOST} 
N.G. Uraltsev, {\em Phys.Lett.} {\bf B376} (1996) 303.  

\bibitem{BSBS} 
M.Voloshin et al., {\em Sov.J.Nucl.Phys.} {\bf 46} (1987) 112.



\bibitem{STONE2} 
I.I.Bigi, B.Blok, M.Shifman, N.Uraltsev, A.Vainshtein, in:  
`$B$ Decays', ed. by S.Stone, World Scientific, 
Rev. Second Edition, 1994, p.132.

\bibitem{POGGIO} 
E. Poggio, H. Quinn, S. Weinberg, {\em Phys.Rev.} {\bf D13} 
(1976) 1958. 

 

\end{thebibliography}
\end{document}